\documentclass{emulateapj}

\usepackage{natbib}
\begin{document}

\title{A serendipitous all sky survey for bright objects in the outer solar system}
\author{M.E. Brown\altaffilmark{1},M.E. Bannister\altaffilmark{2,3},
B.P. Schmidt\altaffilmark{3},
A.J. Drake\altaffilmark{1},S.G. Djorgovski\altaffilmark{1},
M.J. Graham\altaffilmark{1},A. Mahabal\altaffilmark{1},
C. Donalek\altaffilmark{1},S. Larson\altaffilmark{4}, 
E. Christensen\altaffilmark{4}, E. Beshore\altaffilmark{4}, 
R. McNaught\altaffilmark{3}}
\altaffiltext{1}{California Institute of Technology, Pasadena, CA, USA}
\altaffiltext{2}{University of Victoria, Victoria, BC, Canada}
\altaffiltext{3}{The Australian National University, Canberra, Australia}
\altaffiltext{4}{The University of Arizona, Lunar and Planetary Laboratory, Tucson, AZ, USA}
\begin{abstract}

We use seven year's worth of observations from the Catalina Sky Survey
and the Siding Spring Survey
covering most of the northern and southern hemisphere at galactic
latitudes higher than 20 degrees
to search for serendipitously imaged moving objects in
the outer solar system. These slowly moving objects 
would appear as stationary
transients in these fast cadence asteroids surveys, so we develop methods
to discover objects in the outer solar system using individual
observations
spaced by months, rather than spaced by hours, as is typically 
done. While we independently discover 
8 known bright objects in the outer solar system, the faintest having
$V=19.8\pm0.1$, no new objects
are discovered. We find that the survey is nearly 100\% efficient at
detecting objects beyond 25 AU for $V\lesssim 19.1$ ($V\lesssim18.6$ in the
southern hemisphere) and that the probability that there is
one or more remaining outer solar system object of this brightness
left to be discovered in the unsurveyed regions of the galactic
plane is approximately 32\%.
\end{abstract}
\section{Introduction}
The last decade has seen the discovery of most of the brightest objects
in the outer solar system \citep{2003EM&P...92...99T,2008ssbn.book..335B,2009ApJ...694L..45S}.
The wide-field surveys for these 
brightest objects appear moderately complete in both the northern and southern
skies, with only the generally avoided galactic plane and ecliptic poles
left to survey completely.
All surveys miss some fraction of objects ostensibly in their survey regions,
however,
due to temporal gaps, detector gaps,
stellar blending, and numerous other difficulties.
Survey efficiencies have been estimated to be between 
70 and 90\% \citep{2010ApJ...720.1691S, 2011AJ....142...98S,2012AJ....144..140R}, leaving the possibility that 
bright objects in the outer solar system have escaped detection.

While each of the bright objects in the Kuiper belt has yielded an
important boon of scientific information about the origin and evolution
of the Kuiper belt and its objects \citep{2008ssbn.book..335B}, mounting a dedicated survey to 
find this small number of remaining bright objects would be prohibitive.
We note, however, that each of the bright objects discovered over the
past decade serendipitously appeared in multiple other survey images.
These serendipitous detections have been reported to the Minor Planet Center
 in sources from
the {\it Skymorph} data of CCD images from the Palomar 48-inch Schmidt
\footnote{http://skyview.gsfc.nasa.gov/skymorph/} back to 
the original POSS I photographic plates of the 1950s and many in between. 
The objects were
unrecognized at the time of their original imaging owing to the fact
that even at opposition these objects move at speeds of only a few
arcseconds per hour, so they appear identical to stationary stars in
the images.  Only by comparing the images with archival images of the
same location taken at a different time is it recognized that the bright 
outer solar system object appears as a one-time transient.

The large number of serendipitous images of bright outer solar system objects
raises the possibility that a fully serendipitous archival survey could be
attempted to find new objects. An ideal data set for such a survey would
be one which covers large areas of the sky with high enough temporal 
coverage that multiple detection of an object in the Kuiper belt 
could be made. 

Over the past decade, surveys for NEOs have come closest to achieving 
this ideal. In these surveys multiple images are obtained to search for
moving objects, but these images are obtained over a time period too
short to detect the motion of slowly moving objects in the outer solar 
system. Objects in the outer solar system appear simply as stationary
transients.

Here we describe a serendipitous all sky survey for bright objects in the 
outer solar system using archival data from the Catalina Sky Survey (CSS) in
the northern hemisphere and its sister survey, the Siding Spring Survey (SSS), 
in the southern hemisphere.

\section{The CSS, SSS, and CRTS surveys}

The CSS \citep{2003DPS....35.3604L}
operates on the 0.7 m Catalina Schmidt telescope at 
the Catalina Observatory in Arizona and covers 8.1 square degrees
per field to a limiting magnitude of $V\sim 19.5$. 
The CSS has covered approximately 19700 square degrees of sky between
declinations of -25 and +70 at galactic latitudes greater than 10 degrees.
The SSS operates on the 0.5 m Uppsala Schmidt telescope at Siding Spring Observatory in Australia
and covers 4.2 square degrees per field to a limiting magnitude of
approximately 19.0. The SSS has covered approximately 14100 
square degrees of
sky from declination -80 to 0. Accounting for overlap of the two
surveys, the total amount of sky covered is approximately 29700 square degrees.

In both surveys, most fields have been observed multiple times per season
over many years. Figure 1 shows the field coverage with the gray scale 
indicating the number of seasons each field has been observed at
least 4 times. The overall cadence of the surveys is extremely non-uniform; 
the four (or more) observations could be within a single lunation or
could be spread over a six month opposition season. As we will see below, 
however, these details have no effect on our detection scheme. The practical
limit that we find is that an object needs to have been detected at least
4 times in a single season in order for us to extract it from the data set.
\begin{figure}
\epsscale{.8}
\plotone{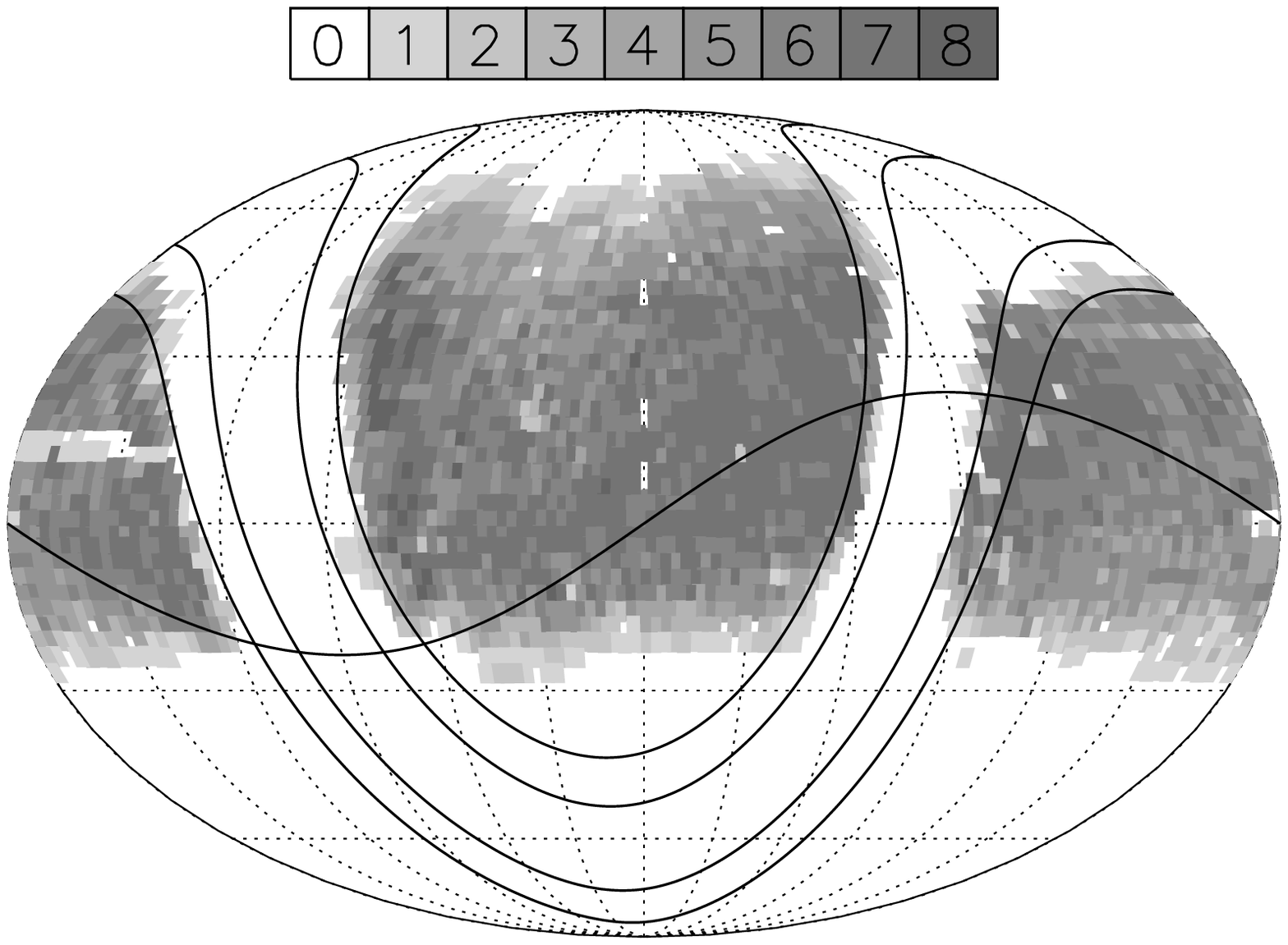}
\plotone{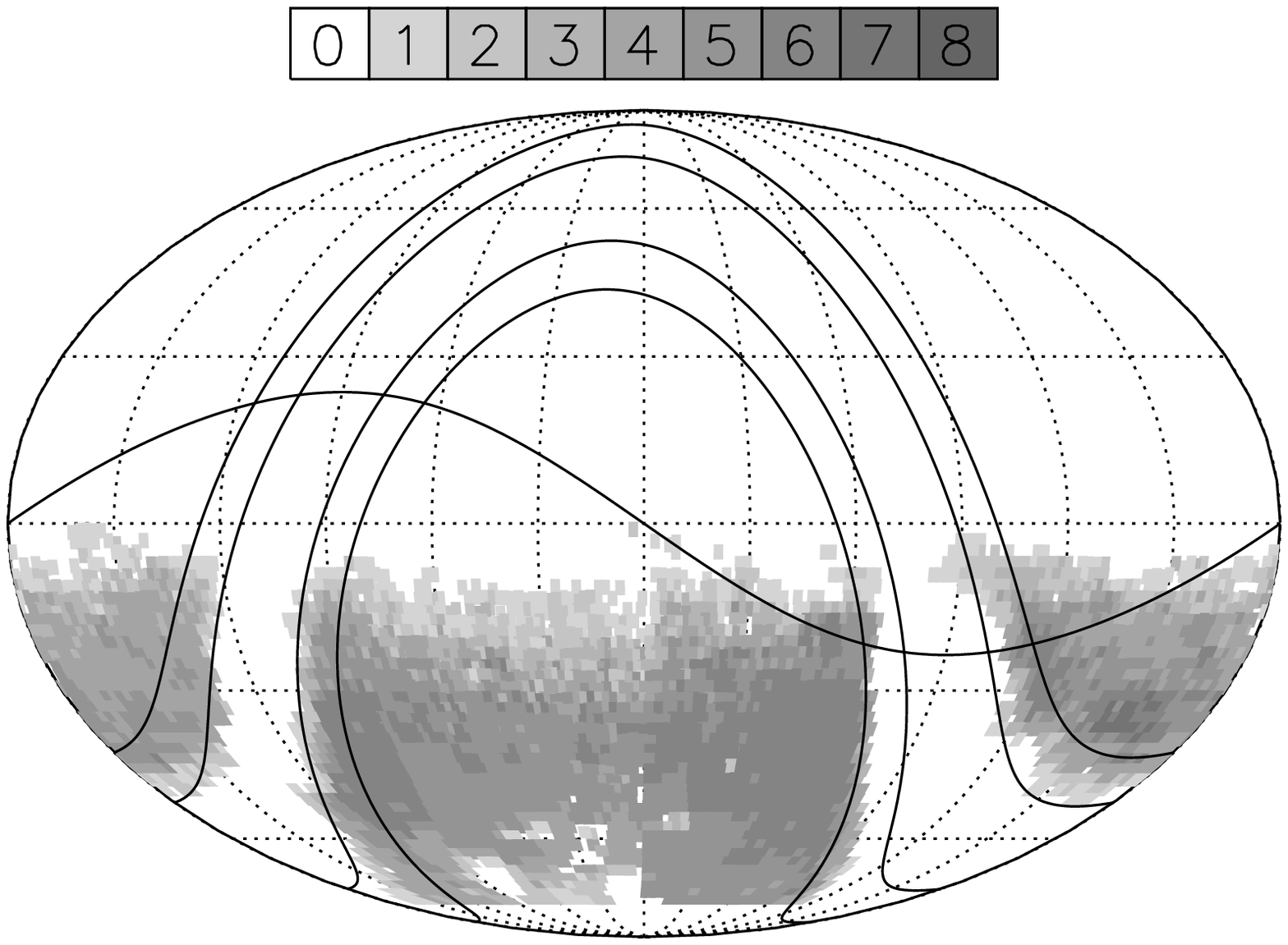}
\caption{(a) Coverage of the Catalina Sky Survey (CSS). The shading shows the number of opposition
seasons in which each field has been covered 4 or more times. The equal area
plot is centered at an RA of 180 degrees and a declination of 0 degrees with 
grid marks placed every 30 degrees of RA and declination. Contours
of galactic latitude of 10 and 20 degrees are shown as well as a line 
showing the ecliptic. (b) Coverage of the Siding Spring Survey (SSS). All parameters are as in (a) except that the plot is centered at an RA of 0 degrees.} 
\end{figure}

In addition to being searched for NEOs and other moving objects,
the CSS and SSS data are analyzed by the Catalina Real-time Transient Survey
(CRTS) \citep{2009ApJ...696..870D} to search for transients. The CRTS compares 
source catalogs from each image with source catalogs generated from
deep coadds of multiple images in addition to comparing with other available
deep catalogs; objects which do not appear in these catalogs are
deemed candidate transients. CRTS performs additional filtering in an
attempt to cull the true astrophysical transients from the large number of 
image artifacts. We begin our analysis, however, with the full catalog
of candidate transients. The full CSS catalog has nearly 1.8 billion
 transient candidates
from 2005 Dec 6 to 2012 April 21, while the SSS catalog has 2.1 billion candidates from 2005 Feb 20 to 2012 April 29.

\section{The slowly moving object search}
\subsection{Creation of the transient list}
The cadence of the NEO surveys has allowed them to detect objects 
out to the orbit of Uranus. 
We will thus define our heliocentric radius of interest
to be 25 AU and beyond. An object in a circular prograde orbit at 25 AU
has a maximum retrograde motion at opposition of 4.9 arcseconds hr$^{-1}$.
Motions higher than this would 
would likely be detected in the NEO surveys.

In typical operations, both the CSS and SSS take four 
images per field per night over a time interval of about 30 minutes.
As our first step in our analysis, we require the detection of
four transient candidates on a single night within a diameter
defined by 4.9 arcseconds times the maximum time interval.
Note that we perform no other filtering here. The four detections
are not required to show linear motion (as they would not for the slowest
moving objects) or have similar measured magnitudes (which they might
not at the magnitude limit of the survey). When four detections
within an appropriate diameter are found, they are collected as
a single transient with the average position, magnitude, and observation
time of the four individual candidate transients. 

Some fraction of the CSS and SSS transients recur at the same location.
These transients are presumably some type of astrophysical source which
has brightness variations sufficiently large that the source does not
appear in the deep catalog but the object occasional becomes bright 
enough to appear in individual images. To remove these clear non-solar
system sources, we search for all transients that have a transient detected
on a different night within 4 arcsecs of the same location. Our final transient list, 
1.2 million sources in the CSS fields and 2.3 million
 sources in the SSS fields, will
contain all outer solar system objects within the geometric
and brightness limits of 
the survey, true astrophysical transients which appear only once at
their location, and image artifacts, which will be the
overwhelming majority of the list.

Figure 2 shows the locations of each of these transients. Significant
structure can be see in the transient locations. In the SSS, in particular, 
transients occur frequently on the field edges, suggesting inconsistent
astrometric solutions in these areas (which will lead stationary stars
to occasionally be classified as transients in large numbers). Similar
effects can be seen for the CSS data in the far north. Other regions
of clear artifact can be seen. In addition, the higher density of transients
in the SSS is clear. The SSS contains 163 transients per square degree
compared to 61 in the CSS. This larger number of southern transients
will make the SSS moving object search comparatively more difficult.
\begin{figure}
\epsscale{.8}
\plotone{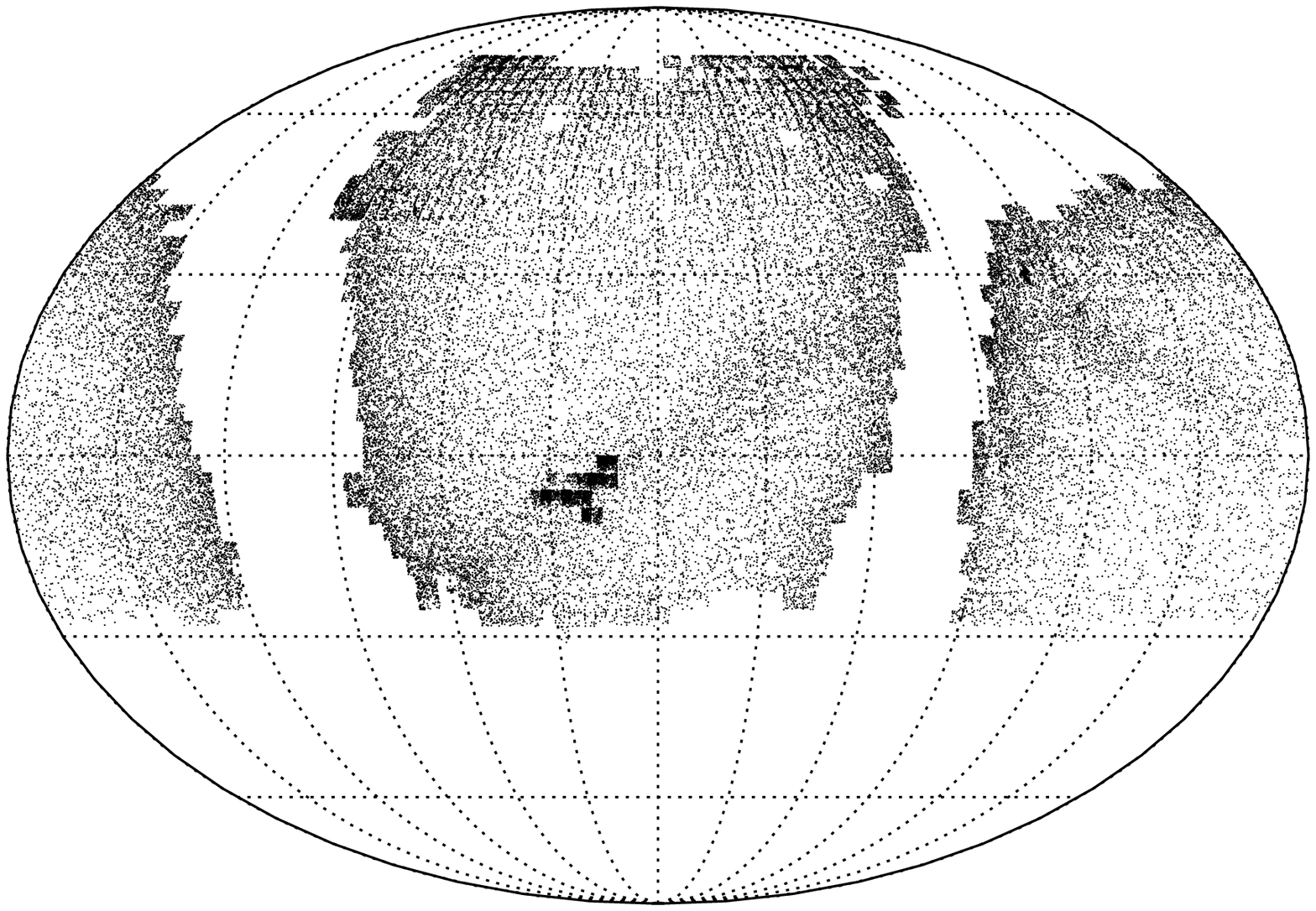}
\plotone{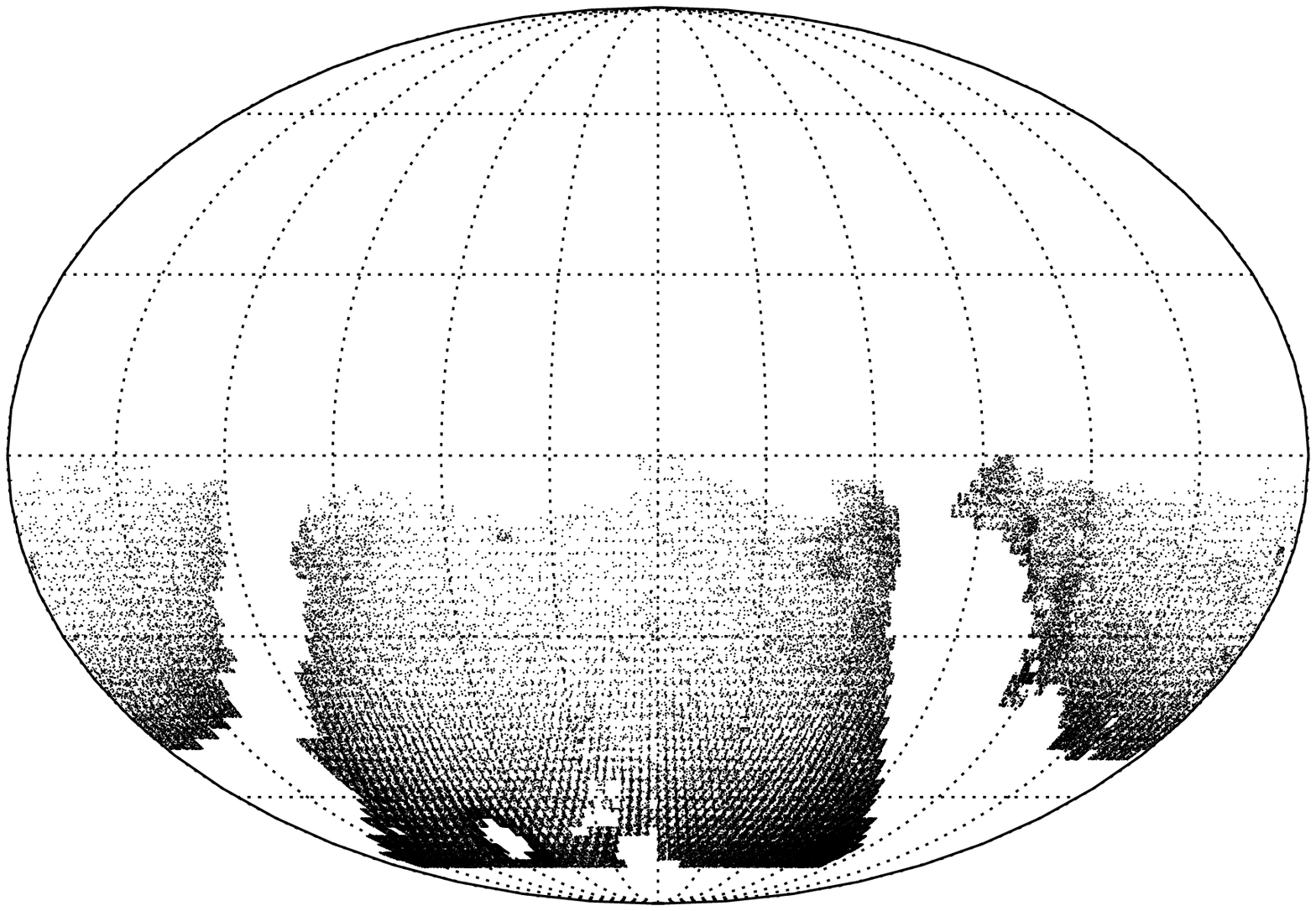}
\caption{(a) The locations of transients in the CSS. To allow viewability only
every 20th transient is shown. Geometry is as in Fig. 1(a). (b) The locations
of transients in the SSS. Every 20th transient is shown. Geometry is as in Fig. 1(b).}
\end{figure}

Nonetheless, real known Kuiper belt objects are also present in the
data. As an example, Figure 3 shows the orbit of Makemake -- the brightest
known KBO after Pluto -- as well as the transients that are detected 
in this region of the sky. Of these 789 transients, 53 are detections
of Makemake itself. The other known bright KBOs likewise have 
many detections.
\begin{figure}
\epsscale{1.}
\plotone{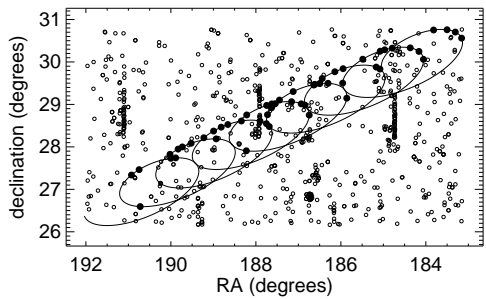}
\caption{All transients detected in the vicinity of Makemake. The path of
the orbit of Makemake over the 8 year period of the survey is shown
as a solid line. The 53 times that Makemake was identified as
a single-night transient are shown as filled circles. Significant
structure is often seen in the transient locations, usually associated
with the edge of a field or with artifacts from bright stars.}
\end{figure}

\subsection{The Keplerian filter}
All objects in the outer solar system move on well-defined Keplerian
orbits. We use this fact to look for collections of transients which
define any physically possible Keplerian orbit. Three points in the sky
at different times are required to define an orbit, but because the three
points contain 9 parameters while an orbit is defined by only
7 parameters, the three points overconstrain the orbit. Thus arbitrary
sets of three points will not be able to be fit to a Keplerian
orbit.

Fitting of orbits is a complex non-linear problem. Attempting to
fit all combinations of three points in the transient list 
($\sim10^{18}$ attempted fits for CSS and $\sim10^{19}$ for SSS) 
is computationally
prohibitive, so we seek methods to minimize the number of
orbital computations required. First, we will consider only combinations 
of transients with pairs of transients separated by no more than 120 days.
This constraint is nearly, though not precisely, equivalent to requiring
three transients observed over a single opposition season. As a short
hand, however, we will refer to this constraint as requiring 3 transients over
a single opposition.

If we confine ourselves to a single opposition season, we can also confine
ourselves to a significantly smaller area of the sky. A stationary object
at 25 AU would have a 5 degree parallax over a six month season. 
We conservatively allow motions up to 5 degrees for our 120 day maximum
separation. 
This single season spatial and temporal filtering
brings the number of potential orbital triplets that need to be
checked down to 160 billion for CSS and 18 trillion for SSS. 
While this initial filter cuts the number of orbits to be fit significantly, 
even this number would be computational prohibitive for full
Keplerian orbital fitting. 

We apply one more filter which is appropriate for these distant objects
observed over a modestly short time interval. As shown in \citet{2000AJ....120.3323B}, 
motions of distant objects in the solar system can be approximated
over a short time period as moving linearly through the solar system
perpendicular to the earth-object vector. This sky plane
approximation is defined by only 6 parameters: the motion vector
in the plane of the sky, the distance to the object, and the position
of the object at a single point in time. For each triplet of transients,
we perform a least-squares fit for these 6 parameters. We then use the
fitted parameters to calculate the predicted positions of the object at
the times of observation and the residuals from this prediction. 

To test the accuracy of this approximation, we examined the positions of
real KBOs over single oppositions. As an example, if 3 observations are made
of Makemake -- at opposition, one month before, and one month after -- the
resulting linear approximation to the orbit predicts the 
position of Makemake to within 25 arcseconds for the four months
surrounding opposition. 
The approximation
is the worst for the the nearest and most eccentric objects. But similar
observations of the positions of 2005 EB299, for
example -- with an eccentricity of 0.51 and a semimajor axis of 52 AU,
and currently near perihelion at 25.7 AU -- predict the position to within
45 arcseconds for the same period.  For our Keplerian filter we 
conservatively require that the maximum 
residual between the linear sky-plane
fit and the data is 50 arcseconds or less. In addition, we require that
the heliocentric 
distance retrieved from the sky-plane approximation be larger than
10 AU and that the orbital energy be within 
a factor of 1.5 of the maximum for a bound object at that distance.
Experimentation with synthetic orbits suggests that these limits
will detect real objects beyond 25 AU in nearly all combinations
of three observations in an opposition season. 
After applying this filter,
we have 140 million triplets remaining in the CSS data and 3 billion in the SSS.

The final step in the Keplerian filter is full orbital fitting.
We use the code of \citet{2000AJ....120.3323B} which efficiently calculates
orbits including planetary perturbations for outer solar system
objects. When the orbits of real KBOs are fit by this code, we
obtain sub-arcsecond residuals. Again, to be conservative, we require
that the orbit of a transient triplet, when fit by this code, 
yield residuals smaller than only 5 arcseconds. In addition we require
that the retrieved heliocentric distance of the object be larger than 15 AU. We
put no other orbital constraints on the fit.

Full Keplerian filtering yields 4.8 million triplets in the CSS data and 235
million 
triplets in the SSS data which can be fit to Keplerian orbits.

%1.17e6 transients in CSS
% 1.6e11 triplets within area
% 1.4e8 triplets through pre-bernstein
% 1192 quadruplets?

\subsection{Further filtering}
Keplerian filtering yields impressive results. For the CSS, for 
example, out of $10^{19}$
possible combinations of three
transients observed across the whole sky over a 7 year period, 
only one out of every two trillion could possibly be fit to single-season
Keplerian orbits for objects
in the solar system beyond 25 AU. Nonetheless, 
these triplets are predominantly not real objects, 
but rather chance alignments of astrophysical transients or noise, so
further filtering is needed.

There are many potential methods we could use to further
filter the transients. We could, for example, require
that all transients in a triplet have the same magnitude within
limits, or we could tighten the constraints on the final
orbit fitting. 
But because of our large data set with the potential that any
real object will be observed many times, we chose instead
to require that there be additional observed transients which also
fit the same orbit.

The simplest requirement to implement is that an orbit be required
to have 4, rather than 3, observations during an opposition season.
The large number of fields in which observations have been made 4 
or more times in a season (Figure 1) suggests that this method could
be quite efficient at detecting objects.

To find quadruplets, rather than just triplets, which fit a Keplerian
orbit, we simply note that any real quadruplets will be made up of combinations
of the triplets that we have already collected. 
We examine all of the triplets that already passed
our initial Keplerian filter and we look for pairs of triplets in
in which two of the three transients in the triplet are common 
across the pair, thus defining 4 unique transients. We pass these
quadruplets of transients through the \citet{2000AJ....120.3323B} orbit fitting routine
and retain any quadruplets for which the calculated $\chi^2$ of the 
fit is below 10.
We find 1192 good quadruplet fits to Keplerian orbits in the CSS
data and 5515 in the SSS data. These numbers are sufficiently small
that we now examine the results in detail.

\section{Results}
\subsection{CSS}
Examining the locations of CSS quadruplets we see that the majority
are tightly clumped into a few distinct locations in the sky (Fig. 4). For
each quadruplet we examine if other quadruplets fit the same orbit by
again running the full orbit fit through the \citep{2000AJ....120.3323B} routine.
In this manner we find that the 1192 CSS quadruplets define 8 distinct
objects (Table 1). 
Each of these recovered objects is, in fact, a known bright object
in the outer solar system. Requiring only 4 detections within an
opposition season reduces the false positive rate in this dataset to zero.
\begin{figure}
\plotone{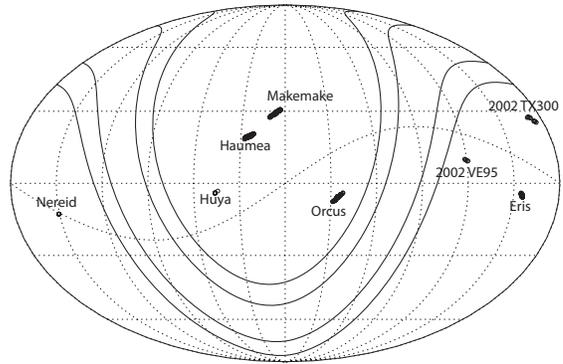}
\caption{The 8 objects detected in our survey. Each of the objects is found
to be a previously discovered object in the outer solar system. The geometry 
is as in Fig. 1(a)}
\end{figure}

While we have no rigorous method of assessing the detection efficiency
of this survey, we estimate the efficiency by examining 
the detections of the brightest known objects. Table 2 shows the
brightest known objects in the solar system beyond 25 AU (with the
exception of Uranus and Neptune, which saturate in our survey).
Looking at the initial unfiltered list of transients, we
determine how many times each object was detected as a transient in
a survey, and, furthermore, how many separate oppositons were linked.
Our algorithm was 100\% efficient at recovering known objects
when 4 or more detections were available in an opposition.

All known outer solar system objects brighter than $V=19.1$ were
detected with only 2 exceptions. Pluto is in a region of the 
galactic plane which was never observed. Quaoar was in the
survey region for only the first season when only 2 images were
obtained. In later seasons Quaoar moved into the galactic avoidance zone. 
Nereid, an irregular satellite of Neptune which does not follow a heliocentric
Keplerian orbit, was detected 14 times and recovered during an opposition season when
its orbit was indistinguishable from a heliocentric orbit.
The bright outer solar system objects with $V\lesssim 19.1$ were all not just detected, they
were observed and would have been detected independently multiple
times. The least well-observed object of this brightness was Orcus.
It was independently detected 381 times as a Keplerian quadruplet. 
It was detected at each opposition in which 4 or more observations were made
of its region.
Any one of these detections would have sufficed. The other bright objects
were independently detected comparable or greater numbers of times. It appears
that for bright objects, where detection is not 
limited by signal-to-noise, our algorithm is essentially 100\% efficient 
at detecting objects when a sufficient number of observations is obtained. 
Because of the nature of the data, which are simply reported detections of
transients, we cannot rigorously prove this assertion of 100\% algorithmic
efficiency through population
modeling, for example, as is often done in surveys. We instead point 
out that the real known objects in the sky provide, essentially, a model
population with thousands of independent quadruplet test cases. Our 
algorithm correctly recovers each of these independent combinations of
real object observations with 100\% efficiency when there are 4 or more
detections in a single opposition season. We conclude that for sufficiently
bright objects in regions with good coverage, we will lose zero objects.

 The 
The next faintest objects detected, 2002 TX300 and Huya,
with predicted $V\sim19.6$ and measured $V\sim19.4$, each have more than a dozen individual detections,
but are only detected 4 times in an opposition
season twice and once, respectively. It is clear that objects of
this magnitude will occasionally be missed not because of algorithmic 
inefficiencies but simply because they will not always be detected even
when they are in the observed field.

Between $19.5<V<20.1$
many objects were detected three or fewer times per opposition season
but only 2002 VE95 (with a measured magnitude of $V=19.8$)
has a single 4-detection opposition. Even with just this single
opposition season in which 2002 VE95 is detected, the algorithm 
correctly identifies 2002 VE95 in the data.

We conclude that the detection efficiency must be nearly 100\% for 
$V\lesssim19.1$ for regions well covered in the survey
and it must begin dropping around $V\sim19.4$ until it 
reaches zero by $V\sim20$. While the precise shape of the efficiency 
cannot be defined, the general shape and behavior appears clear.

\subsection{SSS}
A total of 5515 Keplerian-fitting
quadruplets are found in the SSS survey. None of these quadruplets can
be linked to another quadruplet, thus it appears that these are likely all false
positives. With the significantly higher number density 
of detected transients in the SSS fields, requiring 4 detections
is insufficient for removing all of the false positives. 
Clearly, since none of the quadruplets can be linked, 
adding that requirement that an object be linked 4 times in
each of two opposition seasons reduces the false positive rate to zero.
But we also find that a less stringent requirement -- that we link
the object 5 times in one opposition rather than just 4 times -- is
also sufficient to drop the false positive rate to zero.
It is possible, of course, that one or more of the 5515 quadruplets is
a real object that is sufficiently faint to only have 4 detections
and which is removed by the more stringent filtering. This possibility
demonstrates that the addition of a more stringent detection criterion 
lowers our true detection threshold in this part of the survey. With no 
effective way to perform followup observations of candidates, such a lowering
of the detection threshold is necessary in order to remove
false positives.

Determining an efficiency for the SSS is more difficult with a lack of
detections of real objects, but we make an estimate based on the
experience with the CSS. First, the SSS images are about 0.5 magnitudes less
deep than CSS images. Second, the requirement of 5 detections  raises
the detection threshold. While a 5 detection requirement would
have detected the $V\sim 19.4$ objects in the CSS, all fainter
objects would have been missed. We thus estimate that our detection
efficiency is nearly 100\% for $V\lesssim18.6$ and begins to drop by $V\sim 18.9$.
We have no reliable method of determining where the efficiency drop to zero
but we suspect it happens quickly faintward of $V=18.9$.

\section{Discussion}
No new bright outer solar system objects were detected in this
all sky survey to an approximate limit of $V=19.4$ in the northern
survey and an estimated limit of $V=18.9$ in the southern survey.
If any bright objects remain to be discovered in the outer solar system
they must be at extreme ecliptic latitudes or close to the galactic plane.
No bright objects in the outer solar system have been discovered with
inclinations higher than the 44 degree inclination of Eris, so we do
not anticipate any undiscovered objects at the ecliptic poles. 

We estimate the probability that any bright objects remain to be 
discovered. 
The sky density of bright objects appears approximately uniform within
30 degrees of the ecliptic, while no bright objects have been found
at higher latitudes \citep{2014AJ....147....2S}.
Our survey covered 80\% of the sky within 30 degrees
of the ecliptic. The uncovered 20\% is located within 20 degrees of
the galactic plane.
The survey of \citet{2011AJ....142...98S} covered approximately one third
of the galactic plane below ecliptic latitudes of 30 degrees to a depth
of approximately $R=21$ with a completeness of approximately 75\%.
There are 6 known objects brighter than V$\sim$19, 4 of which were in our
survey region and one recovered in the galactic plane survey.

From these surveys, we estimate that the probability that there is one or more 
remaining objects 
in the outer solar system brighter than $V\sim~19$. To do so we construct Monte
Carlo models of KBO populations with a varying sky densities. Selecting
the simulation which are compatible with the detections of the two 
data sets, we find that the probability that there is one additional
object yet to be detected is 32\%. The probability
that there are two or more is 10\%.
For all regions except for the galactic plane,
these limits extend to
very distant objects; a body at 10,000 AU, for example, would still
move $\sim$10 arcseconds in a month at opposition and would be
detected in our analysis. 

\citet{2014AJ....147....2S} estimate that surveys of the outer solar system
have been approximately 70\% complete to $R=19.5$. This survey suggests that 
at the brightest end the surveys to date have been even more efficient and 
that the most likely scenario is that no new bright objects remain to be 
discovered.

In addition to demonstrating the only modest probability of the existence
of additional bright objects in the outer solar system, this survey 
demonstrates the relative ease of detecting slowly moving solar system
objects in transient surveys.

\acknowledgements This search for moving objects in the CRTS
catalog has been supported by grant
NNX09AB49G from the NASA Planetary Astronomy program.
The CRTS survey was supported by the NSF grants AST-0909182, AST-1313422, 
and AST-1413600.The CSS survey is funded by the National Aeronautics 
and Space Administration under Grant No. NNG05GF22G issued through the 
Science Mission Directorate Near-Earth Objects Observations Program. 
This serendipitous survey was
conceived during a serendipitous conversation at LSST ``All Hands'' meeting
between MEB and MJG.

\eject

\begin{deluxetable}{crrrrrl}
%\rotate
\tablecaption{Measured properties of the detected objects}
\tablehead{ \colhead{linked}    & \colhead{average} & \colhead{a} & \colhead{e} & \colhead{i} & \colhead{distance} & \colhead{identified} \\ 
\colhead{detections}& \colhead{V mag.} & \colhead{(AU)} & & \colhead{(deg)} & \colhead{(AU)} & \colhead{object} }

\startdata
51& 17.06$\pm$.02  & 45.501$\pm$ 0.003&  0.160509$\pm$ 0.00007&  29.002$\pm$ 0.001& 51.867$\pm$.001 & Makemake \\
47& 17.37$\pm$.01  & 43.102$\pm$.002 &  0.195129$\pm$ 0.00009&  28.205$\pm$ 0.001 & 51.217$\pm$.001 & Haumea\\
33& 18.95$\pm$ 0.08 & 39.273$\pm$.002 &  0.22397 $\pm$ 0.00005&  20.567$\pm$ 0.001 & 47.706$\pm$.001 & Orcus \\
28& 18.53$\pm$ 0.01 & 67.82$\pm$ 0.02  &  0.4384$\pm$.0002    &  43.992 $\pm$ 0.001& 96.895$\pm$.006 & Eris \\
 8& 19.5$\pm$ 0.1   & 43.21$\pm$ 0.03  & 0.118 $\pm$ 0.002     & 25.855$\pm$ 0.001  & 41.186$\pm$ 0.004 & 2002 TX300 \\
 4& 19.2$\pm$.1    & 31$\pm$16        &   0.4$\pm$ 0.5        &   2.34$\pm$ 0.06    & 30.37$\pm$.06 & Nereid \\
 4& 19.3$\pm$ 0.1   & 39.74$\pm$ 0.03  &   0.29 $\pm$.01      &  15.49 $\pm$ 0.01  &  28.7$\pm$.01 & Huya \\
 4& 19.8$\pm$ 0.1   & 39.7$\pm$ 0.4    &   0.30$\pm$ 0.01      &  16.33 $\pm$ 0.02  & 28.15$\pm$ 0.02 & 2002 VE95 \\
\enddata
\end{deluxetable}

\begin{deluxetable}{lcccccl}
%\rotate
\tablecaption{The brightest known objects in the outer solar system}
\tablehead{
\colhead{name} & \colhead{V mag} & \colhead{distance} & \colhead{a} & \colhead{e} & \colhead{inc} & \colhead{notes} \\
& \colhead{(avg)} & \colhead{(AU)} & \colhead{(AU)} & & \colhead{(deg)} }
\startdata
134340 Pluto& 14.0& 31.6 & 39.4& 0.25&  17.2& not in survey region       \\
136472 Makemake (2005 FY9)& 16.9& 52.1& 45.45& 0.16& 29.0& {\bf 53 detections, 8 oppositions linked}       \\
136108 Haumea (2003 EL61)& 17.3& 51.1& 43.17& 0.19& 28.2& \bf{49 detections, 6 oppositions linked}       \\
136199 Eris (2003 UB313)& 18.7& 96.7& 67.71& 0.44& 44.2& \bf{39 detections, 6 oppositions linked}       \\
Nereid & 18.8 & 30.0 & 30.1 & 0.01 & 1.76 & \bf{14 detections, 1 opposition linked} \\
50000 Quaoar (2002 LM60)& 18.9& 43.2& 43.58& 0.04&  8.0& not in survey region       \\
90482 Orcus (2004 DW)& 19.1& 47.8& 39.41& 0.22& 20.6& \bf{37 detections, 5 oppositions linked}       \\
55636 (2002 TX300)& 19.6& 41.5& 43.10& 0.12& 25.9& \bf{17 detections, 2 oppositions linked}       \\
28978 Ixion (2001 KX76)& 19.6& 41.7& 39.59& 0.24& 19.6& no detections; low galactic latitude \\
230965 (2004 XA192)& 19.7& 35.9& 46.90& 0.24& 38.1& not in survey area       \\
38628 Huya (2000 EB173)& 19.7& 28.8& 39.77& 0.28& 15.5& \bf{19 detections, 1 opposition linked}       \\
120178 (2003 OP32)& 19.9& 41.4& 43.03& 0.11& 27.2& 1 detection       \\
 (2010 EK139)& 20.0& 39.9& 70.26& 0.54& 29.4& 1 detection in SSS       \\
84922 (2003 VS2)& 20.0& 36.5& 39.29& 0.07& 14.8& 6 detections, max of 3 in single opposition       \\
145451 (2005 RM43)& 20.0& 35.3& 90.35& 0.61& 28.7& 7 detections, max of 3 in single opposition       \\
90568 (2004 GV9)& 20.1& 39.1& 42.17& 0.08& 22.0& too far south for CSS; no detections in SSS       \\
145453 (2005 RR43)& 20.1& 38.6& 43.13& 0.14& 28.5& 6 detections, max of 3 in single opposition       \\
20000 Varuna (2000 WR106)& 20.1& 43.4& 42.91& 0.05& 17.2& 2 detections       \\
47171 (1999 TC36)& 20.1& 30.8& 39.31& 0.22&  8.4& 5 detections, max of 2 in single opposition       \\
145452 (2005 RN43)& 20.1& 40.7& 41.37& 0.02& 19.3& 4 detections, max of 2 in single opposition       \\
229762 (2007 UK126)& 20.1& 45.5& 73.06& 0.49& 23.4& 4 detections, max of 1 in single opposition       \\
55637 (2002 UX25)& 20.2& 41.8& 42.55& 0.14& 19.5& 7 detections, max of 3 in single opposition       \\
278361 (2007 JJ43)& 20.2& 41.8& 48.21& 0.16& 12.1& too far south for CSS; no detections in SSS       \\
55565 (2002 AW197)& 20.3& 46.6& 47.54& 0.13& 24.3& 2 detections       \\
174567 Varda (2003 MW12)& 20.3& 47.9& 45.85& 0.14& 21.5&  1 detection      \\
55638 (2002 VE95)& 20.4& 28.4& 39.22& 0.29& 16.4& \bf{12 detections, 1 opposition linked}       \\
303775 (2005 QU182)& 20.4& 47.9&110.28& 0.67& 14.0& 0 detections       \\
 (2004 NT33)& 20.4& 38.1& 43.41& 0.15& 31.2& not in survey region       \\
202421 (2005 UQ513)& 20.4& 48.8& 43.22& 0.15& 25.7& no detections       \\
144897 (2004 UX10)& 20.5& 38.9& 39.08& 0.04&  9.5& 3 detections, max of 1 in single opposition       \\
119951 (2002 KX14)& 20.5& 39.5& 38.74& 0.04&  0.4& no detections       \\
208996 (2003 AZ84)& 20.5& 45.4& 39.40& 0.18& 13.6& no detections       \\
\enddata
\end{deluxetable}

\end{document}